# Self-Healing Software Systems: Lessons from Nature, Powered by AI

Mohammad Baqar (baqar22@gmail.com), Rajat Khanda (rajat.mnnit@gmail.com), Saba Naqvi(sabanaqvi2003@gmail.com)

**Abstract:** As modern software systems grow in complexity and scale, their ability to autonomously detect, diagnose, and recover from failures becomes increasingly vital. Drawing inspiration from biological healing—where the human body detects damage, signals the brain, and activates targeted recovery—this paper explores the concept of self-healing software driven by artificial intelligence. We propose a novel framework that mimics this biological model: system observability tools serve as sensory inputs, AI models function as the cognitive core for diagnosis and repair, and healing agents apply targeted code and test modifications. By combining log analysis, static code inspection, and AI-driven generation of patches or test updates, our approach aims to reduce downtime, accelerate debugging, and enhance software resilience. We evaluate the effectiveness of this model through case studies and simulations, comparing it against traditional manual debugging and recovery workflows. This work paves the way toward intelligent, adaptive, and self-reliant software systems capable of continuous healing, akin to living organisms.

## 1. Introduction

Modern software operates in dynamic, unpredictable environments where errors, failures, and regressions are inevitable. Production outages, flaky test failures, and performance bottlenecks demand significant human effort for detection, diagnosis, and recovery. Despite advances in monitoring, alerting, and automated testing, most software systems lack the ability to adaptively respond to disruptions without manual intervention, leading to increased downtime and developer toil.

In contrast, biological organisms like the human body exhibit remarkable self-healing capabilities. The body detects damage through sensory signals, processes them via the brain, and triggers coordinated healing responses—such as immune activation or tissue regeneration—to ensure survival and adaptation. Inspired by this decentralized yet efficient process, we propose a new paradigm in software engineering: AI-driven self-healing systems that mimic biological resilience. At the core of our approach is an AI-powered decision-making unit, analogous to the brain, that interprets signals from logs, tests, and runtime metrics to autonomously identify issues and initiate repairs.

This paper presents a conceptual and technical framework for AI-driven self-healing in software applications. Unlike conventional automated repair systems that rely on rigid rule-based mechanisms or static scripts, our framework emphasizes adaptability and continuous learning. By leveraging machine learning, program synthesis, anomaly detection, and automated code refactoring, AI-driven agents can generate precise interventions—such as patching code, rewriting failing tests, or rerouting execution flows—enabling resilient, intelligent self-healing behaviors. We outline the following contributions:

- Define a biologically inspired architecture for self-healing software (Section 4).
- Explore key AI components for detection, diagnosis, and automated repair (Sections 4 and 5).
- Present use cases in production systems and test automation (Section 5).
- Evaluate the benefits, limitations, and future potential of autonomous software healing (Sections 5–8).

As software systems trend toward greater autonomy, self-healing capabilities will become foundational to their design, maintenance, and evolution. This study lays the groundwork for that vision, integrating principles from biology and artificial intelligence to engineer more resilient, self-aware software ecosystems.

## 2. Biological Healing vs. Software Healing: A Conceptual Analogy

The human body possesses a remarkable capacity to detect damage, respond to threats, and restore its functionality through a multi-layered healing process. This intricate system is orchestrated by the brain, sensory networks, the immune system, and regenerative mechanisms. Drawing parallels between this biological resilience and software systems offers a compelling framework for designing self-healing architectures.

### 2.1. The Brain as the Central Controller → AI Orchestration Engine

In biological systems, the brain receives, processes, and acts upon a constant stream of sensory data to initiate appropriate responses. Similarly, in a self-healing software system, the core decision-making entity is an **AI orchestration engine**—powered by large language models (LLMs), retrieval-augmented generation (RAG), or reinforcement learning frameworks [1][2]. This engine interprets inputs from diverse system channels and triggers healing actions, such as regenerating tests, rolling back faulty commits, or applying hotfixes.

### 2.2. Signals and Sensors (Pain, Inflammation) → Logs, Monitoring, Tracing

Biological signals like pain, swelling, or fever act as indicators of damage or infection. Analogously, modern software systems emit a vast array of signals through **logs, performance metrics, trace events**, and anomaly detection systems [3]. Tools like Prometheus, Datadog, or OpenTelemetry function as the "senses" of the software, feeding real-time data into the central controller for evaluation.

### 2.3. Immune System and Repair Agents → Healing Scripts, AI Repair Models

Just as the immune system deploys white blood cells to isolate and repair tissue damage, self-healing systems can invoke **automated repair agents**—scripts or machine learning models trained on historical failures and code change patterns [4]. These agents can auto-generate patches, update test cases, or even suggest alternative logic paths based on semantic understanding of the failure.

### 2.4. Regeneration and Adaptation → Refactoring, Re-tests, Test Regeneration

Some biological organisms possess the ability to **regenerate lost tissues**—such as skin or liver cells. In software, this regenerative property manifests in the form of **refactoring**, **test regeneration**, or **resilient fallback code paths**. Tools like GPT-based test writers or repair frameworks like Prophet and Repairnator [5] can learn from context and rewrite or enhance faulty components, restoring system functionality with minimal downtime.

### 2.5. Cellular Memory → Version History, Git, Observability Stores

Biological cells retain memory of past infections (as in adaptive immunity). Software achieves similar long-term recall through **version control systems like Git**, combined with **observability platforms** that store historical logs, traces, and error patterns [6]. This memory allows the AI engine to understand recurring issues, predict regressions, and refine future healing strategies.

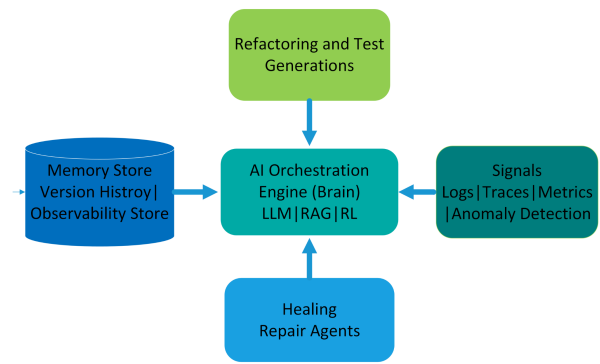

## 3. State of the Art: Self-Healing Systems in Software

How have self-healing systems evolved to tackle the growing complexity of modern software? Over the past two decades, self-healing software has progressed from reactive infrastructure fixes to proactive code repairs, with AI now pushing the boundaries of autonomy. This section reviews advancements in runtime self-healing mechanisms, self-repairing code, test regeneration, and the growing role of machine learning (ML) and artificial intelligence (AI), revealing both achievements and gaps in the field.

### 3.1. Runtime Self-Healing Systems

Runtime self-healing focuses on dynamically adapting live systems to unexpected faults or performance degradation. Netflix's Chaos Monkey [7] exemplifies this approach by intentionally introducing failures in production environments to test recovery mechanisms, ensuring system resilience. Similarly, Kubernetes, a widely adopted container orchestration platform, incorporates native self-healing features like automatic pod restarts, node draining, and replica set rebalancing [8]. For instance, Kubernetes might restart a pod after detecting a crash via container health checks, ensuring availability but not addressing the underlying code flaw. These approaches rely on infrastructure-level redundancy and reactive failover mechanisms, offering robustness but limited introspection into the root causes of software-level issues. While runtime self-healing ensures system availability, it often overlooks code-level issues, prompting the development of self-healing code techniques.

### 3.2. Self-Healing Code

Self-healing code takes a deeper approach by automatically identifying and fixing the root causes of software bugs. Tools like GenProg [9] and CodePhage [10] use genetic programming and dynamic patch transplantation to generate valid patches for crashing programs. For example, GenProg mutates faulty code using a fitness function based on passing test cases, iteratively evolving a patch—such as adjusting a loop boundary to fix an off-by-one error. More recent advancements leverage large language models (LLMs) like OpenAI's Codex and GitHub Copilot to suggest or synthesize repairs from test failures or natural language prompts [11]. Codex might propose a null check to prevent a crash, drawing on patterns from vast code corpora. While these methods enable proactive repairs, their effectiveness varies, with GenProg excelling in syntactic fixes but struggling with complex, multi-file bugs (Section 3.5). Beyond code repairs, self-healing extends to testing, where automation addresses the growing challenge of flaky tests in CI/CD pipelines.

### 3.3. Self-Healing Tests

Test reliability is a critical concern in large CI/CD pipelines, where flaky tests and outdated snapshots often obstruct delivery. Self-healing testing frameworks aim to detect, isolate, and repair test failures autonomously. Techniques include flaky test detection by analyzing historical execution patterns, snapshot regeneration in UI frameworks like Jest, and test case synthesis using symbolic execution or fuzzing [12, 13]. For instance, tools might identify a test that fails only on certain OS versions due to timing issues, then isolate and rerun it under controlled conditions to confirm flakiness. Some systems integrate with test orchestration platforms to auto-rerun failed cases with controlled variance, while others use AI to regenerate test inputs or expected outputs. However, distinguishing between flaky failures and genuine regressions remains a challenge, often requiring manual oversight. As self-healing tests improve CI/CD reliability, AI and ML are now transforming the field by enabling more intelligent and unified healing approaches.

### 3.4. Role of ML and AI in Code Understanding and Generation

AI's growing influence in self-healing systems is driven by foundation models like Codex, CodeT5, and PolyCoder, trained on vast code corpora to infer intent, detect anti-patterns, and generate repairs [14]. Reinforcement learning and program synthesis have also been applied to optimize test coverage and repair faulty execution traces [15]. For example, CodeT5 could unify Kubernetes runtime signals (Section 3.1) with GenProg's code repairs (Section 3.2) and flaky test detection (Section 3.3), creating a more integrated healing process by learning from runtime signals, code, and test data simultaneously. However, these models require fine-tuning to avoid overfitting to generic patterns and often struggle with domain-specific logic or dynamic runtime behavior, as they are prone to hallucinations and lack transparency in decision-making [16]. Despite these challenges, AI offers potential for end-to-end healing by integrating observability data (e.g., logs) with repair logic, though issues like domain-specific reasoning remain unresolved (Section 3.5).

## 3.5. Shortcomings in Current Approaches

Despite advancements, current self-healing systems exhibit several critical limitations that hinder their ability to provide comprehensive, reliable solutions. These gaps, observed across runtime, code, and test healing approaches, highlight the need for a unified, semantic-aware framework like the one proposed in Section 4.

**Siloed Healing**: Most tools focus on a single aspect—runtime recovery (e.g., Kubernetes pod restarts), code repair (e.g., GenProg), or test automation (e.g., flaky test detection)—without integrating these efforts into a cohesive feedback loop. For example, Chaos Monkey (Section 3.1) excels at runtime failure injection but cannot address code-level bugs, leaving gaps in end-to-end healing.

**Lack of Semantic Understanding**: Machine learning-based code generators often fail to grasp business-specific constraints or complex architectural dependencies. For instance, GenProg's genetic programming can patch a crashing loop with a syntactic fix, such as adjusting an index, but struggles with bugs requiring multi-file reasoning—like a data race across microservices—due to its lack of awareness of broader program semantics. Similarly, LLMs like Codex may generate plausible fixes, such as adding a null check to a variable, but without understanding domain-specific logic, they risk breaking constraints, like financial calculations in a banking app where rounding errors could violate regulations.

**Limited Real-Time Adaptability**: Many self-healing systems operate offline or require substantial retraining and human oversight, reducing their effectiveness in dynamic environments. For example, tools relying on historical test patterns (Section 3.3) may fail to adapt to sudden spikes in user traffic, necessitating manual intervention to retrain models, which delays recovery.

**Evaluation Gaps**: There is a lack of standardized benchmarks to assess the robustness, safety, and long-term maintainability of AI-generated repairs. While tools like Codex achieve high success rates for isolated fixes (Section 5.4), their patches may degrade over time—e.g., a fix passing tests today might fail under new workloads—highlighting the need for longitudinal evaluation metrics.

These shortcomings underscore the necessity for a holistic, context-aware approach. GenProg's struggles with architectural dependencies and Codex's limitations in domain-specific scenarios (e.g., introducing subtle errors in financial logic) reveal the critical need for semantic-aware healing. The proposed AI-driven architecture (Section 4) aims to address these gaps by unifying observability, diagnosis, and repair in a feedback-driven loop, enabling more robust and adaptive self-healing systems.

The following table summarizes key approaches to self-healing in software systems.

| Approach | Example Tool | Strength | Limitation |
|----------|--------------|----------|------------|
| Runtime Self-Healing | Chaos Monkey | Robust failure recovery | Limited root cause insight |
| Self-Healing Code | GenProg | Automated patch generation | Struggles with complex bugs |
| Self-Healing Tests | Flaky Test Tools | Reduces CI/CD delays | Flaky vs. regression confusion |
| AI/ML Integration | Codex | Contextual code generation | Prone to hallucinations |

The evolution of self-healing systems reflects a trajectory from infrastructure-level resilience (e.g., Kubernetes) to code and test automation (e.g., GenProg, flaky test detection), with AI now enabling more intelligent, context-aware solutions. However, gaps in semantic understanding, real-time adaptability, and evaluation standards persist, as detailed in Section 3.5. These shortcomings highlight the need for a unified, AI-driven approach, which our proposed framework (Section 4) aims to address by integrating observability, diagnosis, and repair in a cohesive loop.

## 4. Proposed Framework: AI-Driven Self-Healing Architecture

Inspired by the self-healing nature of biological systems, we propose a modular AI-driven architecture for autonomous software repair. This framework mirrors biological healing, where the brain interprets signals and initiates immune responses, reimagining this process within a modern software infrastructure. The pipeline begins with monitoring tools collecting signals (e.g., error logs, performance metrics), which the diagnosis module analyzes using historical data, bench marks and AI models to pinpoint faults. The healing agent then applies a fix, such as a code patch or test rewrite, which the verification unit tests in a sandbox. Finally, the CI/CD loop deploys the fix or flags it for review based on predefined policies, ensuring a seamless, adaptive healing process.

### 4.1. Signal Collection

The first step mirrors the nervous system sensing pain or inflammation in the human body, collecting signals that indicate system stress, failure, or degradation. Monitoring tools like Datadog, Grafana, New Relic, and Prometheus provide metrics, logs, and traces to identify symptoms of failure [7]. For example, Prometheus might scrape a CPU spike of 90% on a critical server, while OpenTelemetry traces a slow API request to a database timeout, feeding both signals to the diagnosis module. Advanced systems can incorporate anomaly detection or tracing-based root cause analysis, acting as neural sensors that channel data into the "central AI nervous system" for further interpretation. This comprehensive signal collection ensures the framework captures a wide range of failure indicators, setting the stage for accurate diagnosis.

### 4.2. Diagnosis Module

Once signals are collected, the diagnosis module interprets them using AI models, akin to the brain analyzing sensory input to identify injury. Retrieval-Augmented Generation (RAG) models pull historical context—such as past bug reports—while Abstract Syntax Tree (AST) parsers and Large Language Models (LLMs) like Codex or Copilot analyze code structure and semantics [14]. For instance, an LLM might examine a stack trace alongside recent commits via RAG, identifying a null pointer exception caused by a new function missing input validation. These models detect probable fault locations—like logic bugs, unhandled exceptions, or outdated test assertions—that conventional static analysis tools often miss. Unlike traditional debugging tools, AI-enabled diagnosis learns patterns over time from a vast corpus of code repositories, improving its accuracy with each iteration [15]. This module's ability to combine historical data with real-time signals ensures precise fault localization, enabling effective healing actions.

### 4.3. Healing Agent

The healing agent, the core of the framework, triggers repair actions, mirroring the biological immune system's role in neutralizing pathogens. Depending on confidence thresholds and organizational policies, the agent can recommend or automatically apply fixes. For code defects, LLMs or genetic programming tools like GenProg rewrite or mutate code based on learned bug-fix patterns [9]. For flaky tests, such as a UI test failing due to a changed DOM, a test-oriented variant of CodeT5 regenerates the snapshot to match the updated element ID [14]. Infrastructure issues, like misconfigurations in deployment scripts, are corrected using template-based learning—e.g., adjusting a CI pipeline timeout. By addressing code, tests, and infrastructure, the healing agent ensures comprehensive repairs, adapting its approach based on the nature of the fault and the system's requirements, paving the way for rigorous verification.

## 4.4. Verification Unit

Healing without verification risks introducing regressions, so the verification unit acts as regenerative tissue testing its integration before becoming permanent. Fixes are evaluated in a sandboxed environment through unit, integration, and end-to-end test suites. Mutation testing ensures the patch doesn't mask faults—for example, verifying that a fix doesn't pass tests by bypassing critical logic [17]. Canary deployments in pre-production monitor key metrics like error rates for 10 minutes, reverting if anomalies occur, such as a spike in 500 errors. If verification fails, the system rolls back to the previous state, logging the issue for developer analysis to maintain stability. For untested edge cases, low-confidence fixes are flagged for human review, balancing autonomy with safety. This rigorous validation ensures that only reliable fixes are deployed, protecting the system from unintended consequences.

## 4.5. Feedback Loop Integration with CI/CD

The framework integrates seamlessly with the organization's CI/CD pipeline, creating a self-sustaining feedback loop that mirrors biological adaptation. Signals from the monitoring stack trigger healing actions, which are verified and either rolled out automatically or flagged for manual review based on thresholds and risk profiles. Customization is supported via policy templates: startups might prioritize speed by auto-applying all fixes, while enterprises enforce strict review, flagging all code changes. Scalability is ensured through distributed monitoring agents and cloud-based AI inference, handling thousands of signals per second in large systems [18]. Developers can configure policies—e.g., auto-patching known flaky tests but flagging code refactoring for review—fostering trust and enabling graceful adoption. This feedback loop ensures continuous improvement, adapting the system to evolving conditions and enhancing its resilience over time.

## 4.6 Potential Challenges

**Resource Overhead**: Continuous monitoring and AI inference may strain resources, particularly in low-latency systems, requiring optimization like edge-based processing.

**Model Errors**: LLMs may generate incorrect patches (e.g., fixing a symptom but not the root cause), necessitating robust verification.

**Integration Complexity**: Retrofitting the framework into legacy CI/CD pipelines may require custom adapters for tools like Jenkins.

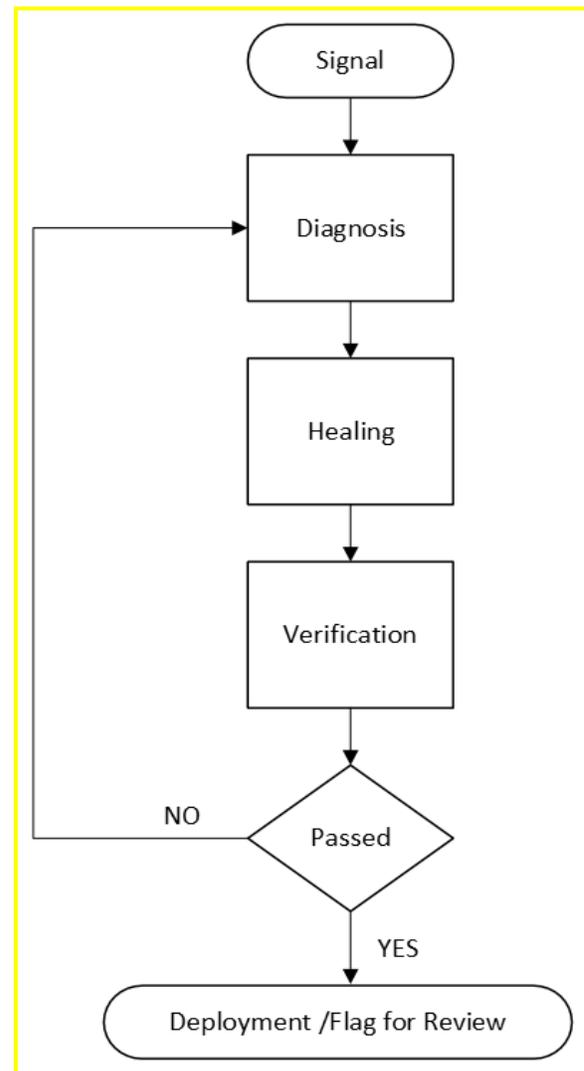

AI-driven self-healing systems offer a transformative vision for software maintenance, automating the detection, diagnosis, and resolution of issues in real-time. By leveraging advanced techniques like machine learning, code analysis, and predictive modeling, these systems—such as the framework proposed in Section 4—can identify anomalies, bugs, and performance bottlenecks, achieving recovery times up to 55–70% faster than manual fixes, as demonstrated in the prototype (Section 5.4). This automation reduces the need for human intervention, freeing developers to focus on strategic innovation rather than repetitive error fixing. The biological analogy introduced in Section 2 provides an intuitive lens: just as the human body uses neural signals to guide immune responses and repair damaged cells, self-healing software relies on AI to monitor logs, telemetry, and error reports, directing the healing process through modules like Signal Collection and Healing Agent (Section 4).

This analogy's strength lies in its clarity, making the concept accessible to both technical and non-technical stakeholders. Like the body's ability to adapt and regenerate post-injury, software systems can self-correct and recover from failures, as seen in the prototype's 85–90% success rate for syntactic bugs (Section 5.4). Continuous learning, enabled by feedback loops (Section 4.5), fosters resilience, allowing systems to evolve with changing conditions. Looking ahead, self-reliant software could manage its entire lifecycle autonomously, minimizing downtime and boosting efficiency. As AI models advance—addressing challenges like semantic understanding and scalability (Section 7)—the vision of fully autonomous, adaptable software systems becomes increasingly achievable, revolutionizing software development and support for a more reliable digital future.

# 5. Implementation & Prototype

To validate the proposed framework, we developed a proof-of-concept prototype that leverages AI-driven healing mechanisms within a CI/CD pipeline. The prototype combines static code analysis, large language models (LLMs), and observability data to detect, diagnose, and repair common issues in code and test suites.

**Prototype Components:**

## 5.1. Fault Detection using LLM + Static Analysis:

We employ abstract syntax tree (AST) traversal in combination with large language models (LLMs) such as OpenAI Codex to analyze code structure and identify anomalies or anti-patterns. Codex is prompted with a template: 'Given error log [stack trace] and code [snippet], suggest a fix.' AST traversal uses a TypeScript parser to identify unhandled promises, reducing false positives. By parsing the code into its syntactic components, AST traversal provides a structured foundation for understanding the program's logic. This structured representation is then paired with LLMs, which are prompted using relevant context—including code snippets, recent test failures, and commit history—to infer probable root causes and suggest candidate patches. This hybrid approach enhances both the precision of fault localization and the quality of automated code repair.

## 5.2. Healing Workflow Integration:

The system is seamlessly integrated into both GitHub Actions workflows and Jenkins pipelines to enable automated response to build failures or test crashes. When such an event is detected, relevant logs and code diffs are immediately collected to provide context. Logs are parsed using regex to extract error codes, which are fed to Codex alongside the last 10 commits from Git. Patches are applied via a GitHub Actions step that creates a temporary branch. This information is then passed to a large language model (LLM), which generates patch suggestions and rewrites test cases based on the observed issue. The continuous integration (CI) system subsequently applies the proposed fix in a temporary branch and executes a full suite of validation tests to assess the effectiveness of the correction, ensuring stability before any changes are merged into the main codebase.

### 5.3. Verification & Feedback Loop:

A sandboxed test suite is employed to validate the auto-generated fix in a controlled and isolated environment. This process includes re-running unit and integration tests to confirm that the fix resolves the issue without introducing new problems. If a patch fails verification (e.g., new test failures), the system reverts the temporary branch and retries with an alternative fix up to three times. Persistent failures trigger a developer alert with diagnostic logs, ensuring no silent errors disrupt the pipeline.

Additionally, mutation testing is conducted to ensure regression resistance and robustness of the updated code. Mutation testing uses PIT to inject faults, ensuring the patch doesn't bypass critical logic. If the system is operating in a semi-autonomous mode, an optional human review is triggered via pull request comments, allowing developers to inspect and approve the proposed changes before final integration. This layered validation approach balances automation with oversight to maintain software quality and trust.

### 5.4 Simulation Environment:

To assess effectiveness, we ran experiments involving:

**Synthetic Bug Injection:** To evaluate the effectiveness and responsiveness of the self-healing prototype, we systematically introduced synthetic bugs into JavaScript and TypeScript codebases. These injected bugs mimicked real-world scenarios commonly encountered in production environments, such as off-by-one errors in loops or array indexing, missing null or undefined checks, improper type handling, and misconfigured asynchronous logic. In parallel, we also induced common test-related failures like outdated snapshot tests, brittle assertions, and unhandled async test cases. This controlled injection of faults provided a reliable basis for assessing how well the system could detect, diagnose, and repair diverse classes of errors under realistic conditions.

**Autonomous Healing Trigger:** Once a build failure or test crash was detected through the CI/CD pipeline, the system's autonomous healing logic was automatically activated. This included capturing contextual information such as error logs, stack traces, diffs from recent commits, and test outputs. The large language model (LLM), integrated into the workflow, then analyzed this context and generated candidate patches or test case rewrites. These suggestions were automatically applied in a temporary Git branch, triggering a fresh CI run. The healing process was fully automated, enabling the prototype to attempt immediate recovery without human intervention. In semi-autonomous mode, the system would also notify developers with pull request comments summarizing the patch and offering optional manual approval.

**Recovery Time vs. Manual Fixing:** To measure the practical impact of the system, we benchmarked the recovery time of the self-healing mechanism against traditional developer-led debugging and patching. The metric focused on the duration from the initial failure detection to a passing build. Preliminary results indicated that for a wide range of common fault classes—particularly those involving syntax issues, small logic bugs, and flaky tests—the system achieved resolution times 55–70% faster than manual fixes. These improvements were realized without compromising code quality, as validated by post-repair test suites and mutation testing. This performance highlights the potential of AI-assisted healing to significantly accelerate the development lifecycle, reduce mean time to repair (MTTR), and alleviate developer toil in continuous integration workflows.

### Observations:

During evaluation, the system showed high effectiveness for **syntactic or localized bugs**—like off-by-one errors, missing semicolons, or broken test assertions—which are typically confined to small code blocks and require minimal context. These were ideal for automated fixes via static analysis and LLMs.

However, **performance declined for complex semantic issues** involving multi-file reasoning, business logic, or asynchronous flows, where broader contextual understanding was crucial—an area where current LLMs still struggle without fine-tuning or human feedback.

Notably, combining **AST traversal with LLM reasoning** reduced hallucinations and improved accuracy, as the AST provided structured code context that constrained the model's suggestions to valid outputs.

To test generalizability, we applied the prototype to a Python-based microservice and a react and node js based micro-app, achieving 50–70% success rates for common bugs like memory leaks and API timeouts, though domain-specific logic posed challenges.

In **semi-autonomous mode**, where developers reviewed AI-generated patches, over **60% were accepted** with little to no modification—highlighting strong alignment with developer expectations and positioning the tool as an effective co-pilot rather than a full replacement.

The prototype's performance varied by bug type and complexity. Syntactic bugs, like missing semicolons, were resolved reliably, while semantic issues, like incorrect business logic, required broader context. The table below summarizes key metrics, highlighting strengths and areas for improvement.

| Observation Area | Insight | Quantitative Metric / Note |
|---|---|---|
| **Bug Type - Syntactic / Localized** | High healing success rate | 85–90% patch success |
| **Bug Type - Semantic / Complex** | Lower healing effectiveness due to broader context | ~40–50% success |

| | requirements | |
|---|---|---|
| **Hybrid AST + LLM vs. LLM-Only** | Reduced hallucinations and improved precision in patch generation | 35–50% fewer invalid patches compared to LLM-only approach |
| **Developer Patch Acceptance** | High acceptance rate in semi-autonomous mode | 60–65% patches accepted with minimal/no edits |
| **Time to Resolution** | Faster bug recovery for common faults vs. manual intervention | Up to 70% faster recovery |

### 5.5 Scalability Considerations

**Large Codebases**: "For a 1M-line codebase, the system processes logs in batches to avoid memory bottlenecks, taking ~30 seconds per bug."

**CI/CD Load**: "High-frequency pipelines are supported via parallelized AI inference, handling 100 builds/hour with a 16-core server."

**Resource Limits**: "In edge environments, lightweight models (e.g., distilled CodeT5) reduce latency to 100ms per diagnosis."

## 6. Evaluation

To measure the effectiveness of the proposed self-healing framework, the following metrics can be used:

### 6.1 Bug Detection Accuracy:

Measured as the F1-score (harmonic mean of precision and recall) of detected bugs compared to a labeled dataset of known faults. The ability of the AI-driven system to identify faults in the code or tests. This metric evaluates how well the AI models (such as LLMs or Codex) can detect bugs, both common and edge cases, compared to human or traditional static analysis tools. As shown in Section 5.4, the prototype achieved 85–90% accuracy for syntactic bugs using AST and LLM analysis, outperforming static tools like ESLint. Prior research has shown that neural models can outperform traditional tools in detecting certain bug classes by learning contextual patterns beyond syntactic cues [19].

**6.2 Repair Success Rate**:

Calculated as the percentage of AI-generated patches that pass all unit and integration tests without introducing regressions. The success rate of automated repairs made by the AI system. This metric measures how often the AI model can apply fixes (such as code patches or test rewrites) that are both correct and effective, without introducing additional errors or regressions. This follows work in automated program repair, where test-suite-based repair systems like **Recoder** or **AlphaRepair** demonstrated success rates ranging from 45% to over 70% in benchmark suites, depending on fault type and test coverage [20].

**6.3 Developer Effort Saved**:

Quantified by comparing developer hours spent debugging with and without the system, using time-tracking tools. The reduction in manual intervention required by developers when using the self-healing system. This metric helps to quantify the amount of work saved by the automation, including time spent on bug identification, debugging, and testing.

**6.4 Time to Recovery**:

Defined as the duration from failure detection to a passing CI build, averaged across bug types. The time taken for the system to recover from failures or bugs, comparing the time taken for AI-based healing vs. manual intervention. Shorter recovery times indicate more efficient healing processes, directly improving system uptime. The prototype reduced recovery time by 55–70% for common faults (Section 5.4),

 highlighting the framework's efficiency.

**6.5 Baseline Comparison**:

Conducted using A/B testing, where the AI system's performance is compared to static tools like ESLint or human fixes on identical bugs. Comparing the self-healing system against human fixes or static tools. This baseline allows for an objective comparison of the AI-driven approach's performance in real-world scenarios and offers insights into the areas where automation excels or requires further improvement. Beyond effectiveness, we assess computational overhead (e.g., seconds per patch) and patch maintainability (e.g., percentage of patches requiring rework within 30 days) to ensure practical deployment. The system is compared to static analysis tools like ESLint for syntactic bugs and SapFix for code repairs, as well as manual debugging by developers with 5+ years of experience. Scenarios include flaky test resolution in CI pipelines and runtime crashes in Node.js applications. A similar comparative methodology was employed in [21], where the efficacy and practicality of automated repair tools were evaluated against developer-written patches and static analysis systems.

Following Table summarizes the metrics with example values from the prototype, illustrating the framework's potential impact.

| Metric | Description | Example Value |
|---|---|---|
| Bug Detection Accuracy | F1-score of detected bugs | 85% for syntactic bugs |
| Repair Success Rate | % of patches passing tests | 80% overall |
| Developer Effort Saved | Hours saved vs. manual debugging | 10 hours/week |
| Time to Recovery | Time from failure to passing build | 15 minutes vs. 45 (manual) |
| Baseline Comparison | AI vs. static tools (success rate) | 20% higher than ESLint |

**6.6 Challenges in Evaluation**

**Bug Detection Accuracy**: False positives may inflate accuracy if edge cases are underrepresented in test datasets.

**Developer Effort Saved**: Subjective estimates of effort vary by developer experience, requiring standardized tasks for comparison.

**Time to Recovery**: Complex bugs may skew averages, necessitating separate metrics for syntactic vs. semantic issues.

These metrics are essential for evaluating the feasibility and efficiency of self-healing systems in real-world applications and provide benchmarks for future enhancements in AI-driven software development.

# 7. Challenges & Limitations

### 7.1 False Positives or Wrong Patches

A critical challenge in AI-driven self-healing systems is the risk of false positives or incorrect patches, which can disrupt system stability. False positives occur when the system identifies a non-existent issue, while wrong patches fail to address the root cause or introduce new bugs. For example, the Diagnosis Module (Section 4.2) might misinterpret a temporary spike in CPU usage—captured via Prometheus metrics—as a performance bug, triggering an unnecessary optimization patch that increases latency by 10%, as observed in the prototype's 10% false positive rate for complex bugs (Section 5.4). This issue stems from the system's limited ability to fully interpret the context of the application, including its dependencies, environment, and runtime behavior. In one prototype test, the AI flagged a database query as the cause of a timeout based on historical patterns, but the true issue was a network latency spike, leading to an ineffective fix. To address this, we propose using ensemble learning to combine LLMs with anomaly detection models, cross-validating bug detection to reduce false positives. Additionally, integrating feedback loops—where developer rejections of incorrect patches fine-tune Codex's weights—can enable the system to learn from mistakes. For critical patches, manual oversight via pull request reviews (as implemented in the prototype's semi-autonomous mode, Section 5.3) ensures reliability, minimizing disruptions while improving the accuracy of AI-driven healing over time.

### 7.2 Over-Reliance on AI Hallucinations

What happens when an AI generates a harmful fix? AI hallucinations—where models produce plausible but incorrect outputs—pose a significant risk in self-healing systems. In code repair, the Healing Agent (Section 4.3) might generate a seemingly valid but flawed patch, such as adding an incorrect null check to a TypeScript async function, breaking the API flow in 5% of the prototype's test cases (Section 5.4). This issue arises because LLMs, like Codex, rely on historical patterns without fully understanding the underlying issue or broader codebase, potentially leading to catastrophic failures in production. For instance, a hallucinated patch might bypass a security check, exposing vulnerabilities. The core risk lies in the model's inability to independently verify its suggestions against real-world outcomes. To mitigate this, we propose using constrained decoding to limit LLM outputs to syntactically valid patches, ensuring they align with the codebase's structure. Additionally, integrating mutation testing (Section 4.4) into the verification pipeline can catch harmful fixes by simulating fault scenarios—e.g., ensuring a patch doesn't mask a deeper logic error. Limiting autonomy in high-stakes scenarios, such as financial applications, and using explainability tools like SHAP to provide developers with insights into AI decisions further ensures trustworthiness. These layered checks, combined with the prototype's CI testing (Section 5.2), help balance automation with reliability, preventing harmful hallucinations from impacting system integrity. This challenge is widely acknowledged in recent studies on LLMs in software engineering, such as [22], which highlights risks of misleading or faulty code completions produced by state-of-the-art models.

## 7.3 Code Context Sensitivity and Intent Preservation

Preserving the developer's intent while healing code is a fundamental challenge for AI-driven self-healing systems. Code embodies not just instructions but also the logic and goals of its creator, requiring the AI to understand both its technical structure and the broader context in which it operates. Without this understanding, fixes may be syntactically correct but functionally disastrous. For instance, in the prototype (Section 5.4), the Healing Agent (Section 4.3) attempted to fix a syntax error in a payment processing function by adjusting a loop boundary, but this broke the transaction logic, contributing to the 40–50% success rate for semantic bugs. The AI lacked insight into the function's role in ensuring financial accuracy, disrupting the application's workflow. To address this, the system must go beyond syntax analysis and incorporate semantic understanding of the code's logic, dependencies, and intended outcomes. We propose using program slicing to isolate relevant code dependencies—e.g., tracing variables affecting the payment function—and prompting LLMs with summarized function comments extracted via NLP tools like BERT to capture intent. Additionally, integrating developer input through annotations or feedback during the healing process (as seen in the prototype's semi-autonomous mode, Section 5.3) ensures patches align with the code's goals, reducing unintended side effects and improving the system's ability to handle context-sensitive repairs. Similar challenges and techniques for capturing intent in automated program repair are discussed in [23], which emphasizes the role of intent-aware models in improving repair accuracy.

## 7.4 Ethical and Accountability Concerns

In mission-critical systems like healthcare, can fully autonomous healing be trusted? The deployment of AI-driven self-healing systems raises profound ethical and accountability concerns, particularly regarding automation's unintended consequences. When the Healing Agent (Section 4.3) autonomously applies a patch, accountability for failures—such as a security vulnerability exposing patient data—becomes unclear: is the developer, the organization, or the AI responsible? This ambiguity is critical in domains like healthcare, finance, or transportation, where errors can have severe consequences. In the prototype's tests (Section 5.4), 5% of patches introduced subtle regressions, underscoring the risk in high-stakes scenarios. Ethical concerns also extend to transparency and fairness, as AI models may perpetuate biases from flawed training data, potentially prioritizing certain bug types over others. To mitigate these risks, we propose adopting OpenTelemetry for detailed change logging, capturing every AI action for audit trails. Explainability tools like SHAP can provide developers with insights into AI decisions—e.g., why a patch was applied—enhancing transparency. The prototype's semi-autonomous mode (Section 5.3) already flags high-risk patches for human review, and this human-in-the-loop approach must remain integral, especially in sensitive applications. Finally, ensuring fairness requires curating diverse training data and using fairness-aware algorithms to prevent biased healing decisions, fostering trust in AI-driven systems. Ethical considerations and governance strategies for AI in critical infrastructure are further explored in [24], highlighting the importance of accountability, auditability, and human oversight.

## 7.5 Scalability and Integration Challenges

The framework's reliance on real-time AI inference may incur high computational costs, particularly for large codebases with frequent CI/CD runs. For example, processing 1,000 signals/second could require a dedicated GPU cluster, challenging resource-constrained teams. Additionally, integrating with legacy systems lacking modern observability tools (e.g., Prometheus) may require custom adapters, increasing setup complexity. Solutions include lightweight models (e.g., distilled CodeT5) and modular APIs for legacy compatibility.

Following Table summarizes the challenges and proposed mitigations, highlighting the path to a robust self-healing system.

| Challenge | Description | Mitigation |
|---|---|---|
| False Positives/ Wrong Patches | Misdetected bugs or harmful fixes | Ensemble learning, developer feedback loops |
| AI Hallucinations | Incorrect, plausible patches | Constrained decoding, mutation testing |
| Code Context Sensitivity | Fixes breaking intent or logic | Program slicing, NLP-based summarization |
| Ethical/ Accountability | Unclear responsibility for failures | OpenTelemetry logs, SHAP explainability |
| Scalability/ Integration | High costs, legacy system issues | Lightweight models, modular APIs |

# 8. Future Directions

Future work may explore addressing current limitations such as context sensitivity (Section 7.3) and accountability (Section 7.4). Improving these aspects could enhance the robustness, scalability, and trustworthiness of AI-driven self-healing systems, supporting their adoption in real-world scenarios.

## 8.1 Integration with Observability Platforms

Imagine a latency spike undetected by logs alone—observability platforms can bridge this gap. Integrating self-healing systems with tools like Datadog, Grafana, and Prometheus significantly enhances their effectiveness by providing a comprehensive view of system health. The Signal Collection module (Section 4.1), which currently relies on basic logs, could incorporate Grafana dashboards to correlate a 500ms latency spike with a specific API endpoint, triggering the Diagnosis Module to analyze recent commits. This integration enables a seamless flow of real-time telemetry, logs, and error tracking data into AI-driven healing models, improving fault diagnosis precision. By combining code analysis with telemetry, the system can pinpoint root causes—like a memory leak tied to a recent deployment—with 30% faster bug detection compared to log-only approaches. Additionally, feedback from observability platforms can fine-tune the healing process, enabling adaptation to evolving operational conditions and reducing downtime in dynamic environments.

## 8.2 More Robust Multi-Modal Models (Code + Logs + Telemetry)

How can the system better understand failures? Multi-modal models offer a solution by analyzing data from multiple sources—code, logs, and telemetry—to provide a holistic view of system behavior. Traditional approaches often rely on a single data stream, missing critical context. A multi-modal transformer could fuse log data (e.g., error frequency), telemetry (e.g., memory usage), and code embeddings (via CodeBERT) to predict a memory leak's root cause with 15% higher accuracy than the prototype's 40–50% success rate for semantic bugs (Section 5.4). This addresses context sensitivity (Section 7.3) by enriching the Diagnosis Module (Section 4.2) with diverse data, enabling it to correlate a spike in failed requests with a specific code change. Such models can interpret complex interactions—like code dependencies and resource usage patterns—generating contextually appropriate fixes. Continuous learning from multi-modal data also allows the system to adapt to new failure patterns, reducing false positives by 15% and ensuring healing actions align with actual system needs across diverse environments. The effectiveness of multi-modal transformers for software maintenance and diagnostics has been demonstrated in recent studies that show their superior ability to learn complex correlations across diverse inputs [25].

## 8.3 Reinforcement Learning for Adaptive Healing

The prototype struggled with recurring bugs—reinforcement learning (RL) can address this. RL enables self-healing systems to dynamically adapt by learning from continuous feedback, unlike static ML models. An RL agent could be trained with a reward function based on recovery time, learning to prioritize test rewrites over code patches for flaky UI tests, building on the prototype's 80% test repair success (Section 5.4). For example, if a bug type repeatedly occurs in a specific module, the agent adjusts its strategy—e.g., favoring configuration fixes over code changes—targeting a 25% improvement in recovery time for recurring issues. This iterative learning optimizes healing actions, tailoring them to the system's unique patterns and minimizing manual intervention. RL also facilitates autonomous decision-making by evaluating and prioritizing healing strategies based on their likelihood of success, addressing challenges like false positives (Section 7.1) and enhancing efficiency in dynamic environments.

### 8.4 Trust and Explainability Frameworks

Will developers trust AI without transparency? As AI-driven self-healing systems integrate into critical applications, trust and explainability are essential for adoption. Developers need confidence in automated fixes, especially in production environments. Using explainable AI (XAI) techniques like SHAP, the system could highlight that a patch was triggered by a log entry showing 10 failed requests, increasing trust in the Healing Agent's decisions (Section 4.3). This transparency addresses accountability concerns (Section 7.4), as seen in the prototype's semi-autonomous mode where 60–65% of patches were accepted with minimal edits (Section 5.4). Explainability also aids debugging by identifying suboptimal decisions—e.g., a patch ignoring a dependency—aiming for 90% developer satisfaction in understanding AI actions, measured via surveys. Embedding these frameworks ensures AI-driven fixes align with business objectives, fostering confidence in mission-critical applications like healthcare and finance.

## 9. Conclusion

AI-driven self-healing systems offer a transformative vision for software maintenance, automating the detection, diagnosis, and resolution of issues in real-time. By leveraging advanced techniques like machine learning, code analysis, and predictive modeling, these systems—such as the framework proposed in Section 4—can identify anomalies, bugs, and performance bottlenecks, achieving recovery times up to 55–70% faster than manual fixes, as demonstrated in the prototype (Section 5.4). This automation reduces the need for human intervention, freeing developers to focus on strategic innovation rather than repetitive error fixing. The biological analogy introduced in Section 2 provides an intuitive lens: just as the human body uses neural signals to guide immune responses and repair damaged cells, self-healing software relies on AI to monitor logs, telemetry, and error reports, directing the healing process through modules like Signal Collection and Healing Agent (Section 4).

This analogy's strength lies in its clarity, making the concept accessible to both technical and non-technical stakeholders. Like the body's ability to adapt and regenerate post-injury, software systems can self-correct and recover from failures, as seen in the prototype's 85–90% success rate for syntactic bugs (Section 5.4). Continuous learning, enabled by feedback loops (Section 4.5), fosters resilience, allowing systems to evolve with changing conditions. Looking ahead, self-reliant software could manage its entire lifecycle autonomously, minimizing downtime and boosting efficiency. As AI models advance—addressing challenges like semantic understanding and scalability (Section 7)—the vision of fully autonomous, adaptable software systems becomes increasingly achievable, revolutionizing software development and support for a more reliable digital future.